# Split-by-edges trees

Asbjørn Brændeland

*Abstract*

A *split-by-edges* tree of a graph $G$ on $n$ vertices is a binary tree $T$ where the root = $V(G)$, every leaf is an independent set in $G$, and for every other node $N$ in $T$ with children $L$ and $R$ there is a pair of vertices $\{u, v\}$ ⊆ $N$ such that $L = N - v$, $R = N - u$, and $uv$ is an edge in $G$. The distance from the root to an independent set $I$ is $n - |I|$ and the maximum independent sets of $G$ are the ones closest to the root. For every independent set $X$ of $G$ there is a leaf $Y$ in $T$ such that $X \subseteq Y$, thus every maximal independent set in $G$ is a leaf in $T$. In a uniquified split-by-edges tree a maximum independent set is found in a layer-by-layer search in at the most $2^{\phi(n)}$ time, in terms of number of split operations, with $\phi(n) = O(0.369425n)$ for random graphs.

## The SBE-tree

An *independent set $I$* in a graph $G = (V, E)$ is a set of vertices no two of which are adjacent. If $I$ is not a proper subset of another independent set in $G$ then $I$ is a *maximal independent set*, and if $G$ has no larger independent set then $I$ is a *maximum independent set* of $G$. Tarjan and Trojanowski point out that for every vertex $v \in V$, any maximum independent set of $G$ must be a subset of either $V - v$ or $V - N(v)$ (when $N(v)$ are the neighbors of $v$), using that as the starting point for an algorithm that finds a maximum independent set in less than $2^n$ time [1]. Relatedly, given an edge $uv \in E$ and a maximum independent set $M \subset V$, either $M \subseteq V - u$ or $M \subseteq V - v$. This gives rise to the following definition.

**Definition 1**: Let $G$ be a graph and let $T$ be a binary tree of subsets of $V(G)$. Then $T$ is a **split-by-edges tree**, or **SBE-tree**, of $G$ if and only if the root of $T = V(G)$, every leaf in $T$ is an independent set of $G$, and for every other node $N$ in $T$ with children $L$ and $R$ there is a pair of vertices $\{u, v\} \subset N$ such that $L = N - u$, $R = N - v$, and $u$ and $v$ are adjacent in $G$.

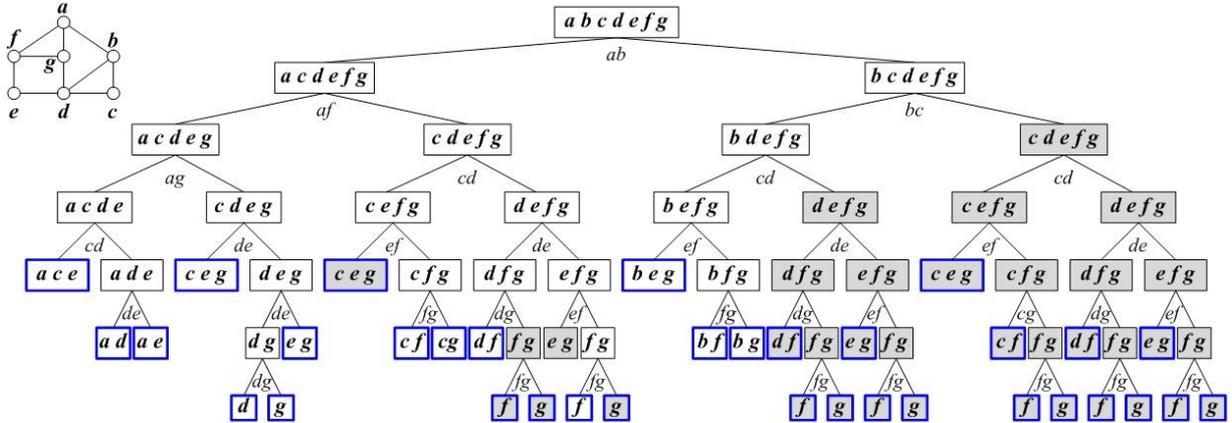

Figure 1.  An SBE-tree of the graph at the upper left. The leaves have bold blue frames. The gray nodes are duplicates of others. The branching labels, which do not belong to the tree, show the splitting edges.

**Theorem 1**. *Given a graph $G$ and a split-by-edges tree $T$ of $G$, for every independent set $X$ of $G$ there is a leaf $Y$ in $T$ such that $X \subseteq Y$.*

*Proof*: Given an independent set $I$ of $G$, for every node $N$ with children $L$ and $R$ in $T$, if $I \subset N$ then $I \subseteq L$ or $I \subseteq R$.  □

**Corollary 1.1**. *Every maximal independent set of $G$ is a leaf in $T$.*  □

The number of possible child pairs of an SBE-tree node equals the number of neighbor pairs in the node. If $G$ is the *n*-complete graph a *k*-vertex node in the SBE-tree of $G$ contains $\binom{k}{2}$ neighbor pairs, the number of possible quadruples of grandchildren is $\binom{k}{2}\binom{k-1}{2}^2$, the number of octuplets of grand-grandchildren is $\binom{k}{2}[\binom{k-1}{2}\binom{k-2}{2}^2]^2$, etc. E.g. $K_6$ has $15(10(6 \cdot 3^2)^2)^2 = 12{,}754{,}584{,}000$ SBE-trees. However, by Theorem 1, for each $l$, the contents of layer $l$ is the same in all of these trees.

If $G$ is not complete, the contents, shapes and sizes of its SBE-trees vary, dependent on the order in which the edges are searched, as illustrated in Figure 2. But notice that for every layer $L_l$ down to the one that contains the maximum independent sets, $|L_l| = 2^l$.

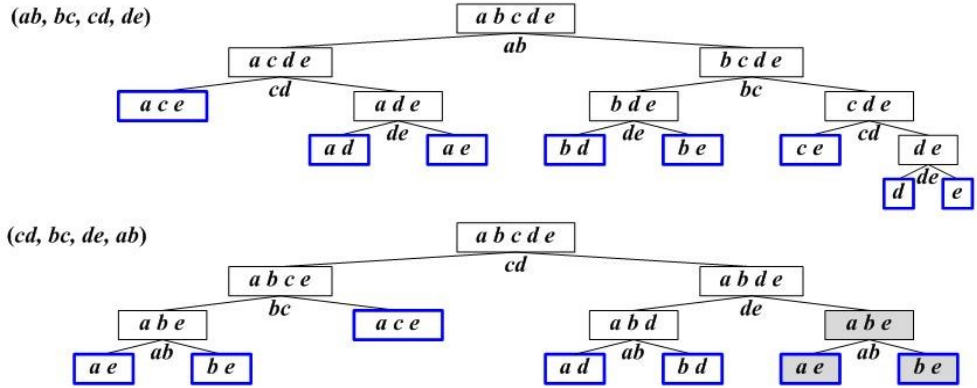

Figure 2. The figure shows two SBE-trees of the path *abcde*. The edge search order is given above and to the left of each tree. Both trees satisfy Definition 1, but their contents, shapes and sizes are different.

**Definition 2**: The **ordered SBE-tree** of a graph $G$ is the SBE-tree of $G$ in which the splitting edges have been selected in ascending order, when $uv < wx$ if an only if $u < w \lor (u = w \land v < x)$.

By the definite form *'the SBE-tree of G'* is always meant *'the ordered SBE-tree of G'*.

The SBE-tree $T$ of a graph $G$ can be generated from any single node $N$ in $T$. I.e., if $N$ is not the root of $T$ there must be an edge $uv$ in $G$, found in a reverse order, such that $u \in N$ and $v \notin N$, and then $N + v$ is the parent and $N + v - u$ is the sibling of $N$ in $T$.

*The uniquified SBE-tree*

For most graphs, every SBE-tree contains duplicate vertex sets, and the tree size can be many times the cardinality of the corresponding set.

**Definition 3**: An SBE-tree minus its duplicate nodes is a **uniquified SBE-tree**, or a **USBE-tree**.

Let $T$ be the SBE-tree and $T'$ the corresponding USBE-tree of $G$. If $G$ is a complete graph on $n$ vertices the size of $T$ is $2^n - 1$ and the size of $T'$ is $\binom{n+1}{2}$. If $G$ is not complete, the exclusion of duplicates has less effect, but this is to some extent outweighed by the occurrence of leaf nodes closer to the root, i.e., for each non-singleton leaf $\phi$ in $T$, the size of $T$ is reduced by $2^{|\phi|} - 1$ compared to the SBE-tree of a complete graph of the same order as $G$. Below are SBE and USBE-tree sizes, etc., for some graphs.

| $G$ | $n$ | $m$ | $\delta$ | $\Delta$ | $\alpha$ | $|SBE(G)|$ | $|USBE(G)|$ |
|---|---|---|---|---|---|---|---|
| $K_{48}$ | 48 | 1128 | 47 | 47 | 1 | 281474976710655 | 1176 |
| 18-regular graph | 48 | 432 | 18 | 18 | 6 | 205624938644223 | 192146 |
| Apollonian network | 48 | 138 | 3 | 6 | 12 | 54263808384247 | 17721342 |
| Möbius ladder | 48 | 72 | 3 | 3 | 23 | 238972941719 | 153349985 |
| Path | 48 | 47 | 1 | 2 | 24 | 15557484097 | 15557484097 |

The SBE-tree of an edgeless graph has trivially (and by definition) just a single node. For connected graphs, the extreme cases, the complete graph, and the path, are determined as such by their density (disregarding heavily leafed graphs, such as stars, which rapidly break down to edgeless graphs).

In the USBE-tree of a complete graph, the width of layer $l$ is $l$, and every independent set is in the bottom layer. In the USBE-tree of a path on $n = 2r$ vertices, the widths of the first $r + 1$ layers are $2^l$, and the combined size of these layers is $2^r + r$. The tree width continues to grow, at a slowing rate, until it reaches a maximum somewhere before three quarters from the root.

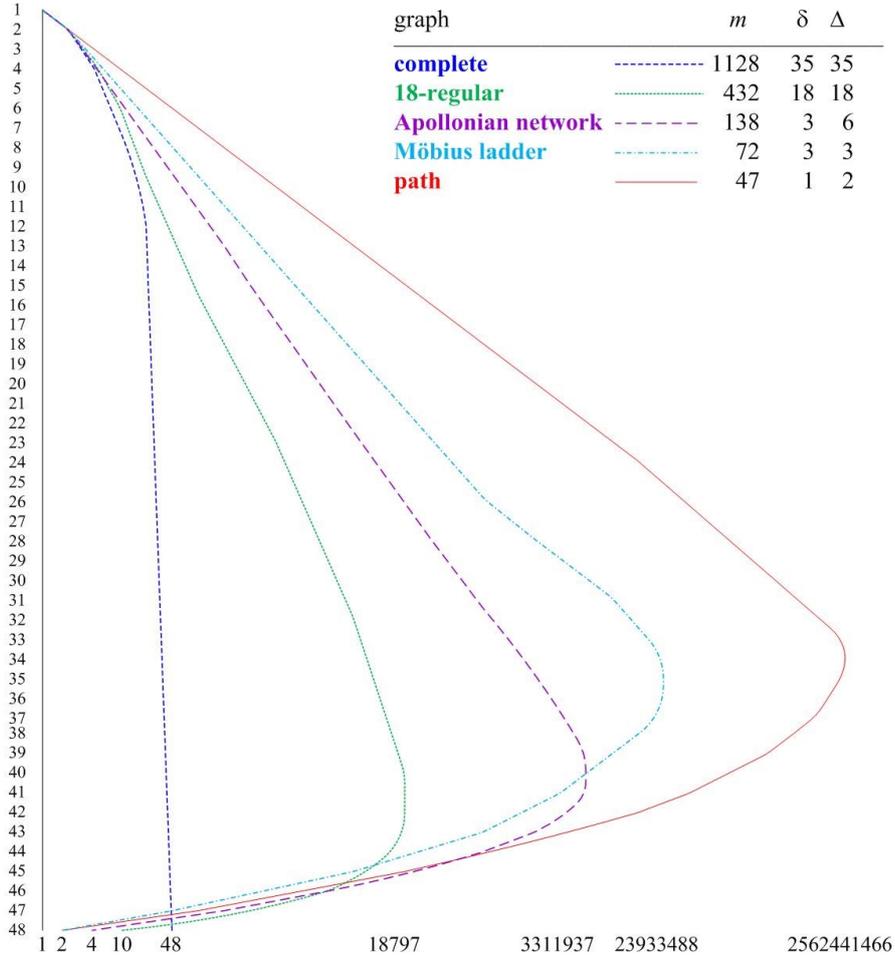

Figure 3. USBE-tree layer widths for 5 graphs on 48 vertices. The horizontal scale is $2 \log_2 w$ (= layer width).

**Claim 2.** *The SBE-tree of a path contains no duplicates.*

*Proof*: Let $P = (\{p_1, \ldots, p_n\}, \{p_1 p_2, \ldots, p_{n-1}, p_n\})$, let $T$ be the SBE-tree of $P$, let $N = (p_i, \ldots, u, \ldots, v, \ldots)$ be a node in $T$ such that $\{u, v\}$ splits $N$ into $L$ and $R$, and let $I = (p_i, \ldots, u)$. Since the splitting edges have been selected in ascending order, $I$ is an independent set in $P$ and must occur in every node below $L$ and, since $u$ is not in $R$, $I$ cannot occur in any node below $R$, thus the SBE-tree of $R$ cannot contain a duplicate of any node in the SBE-tree of $L$. □

The cardinality of a maximum independent set of an $n$-vertex path $P = \lfloor (n + 1)/2 \rfloor$ and in the SBE-tree of $P$, the MI sets are in layer number $\lfloor n/2 \rfloor + 1$.

By Claim 2 every split in the SBE-tree of a path gives a pair of unique nodes. This makes it possible to compute the layer widths of the tree in linear time, without constructing any part of the tree.

## USBE-trees of random graphs

The graphs giving the numbers and curves in Figure 3 are structurally clear and well suited to illustrate the relation between graph density and USBE tree layer widths in general.

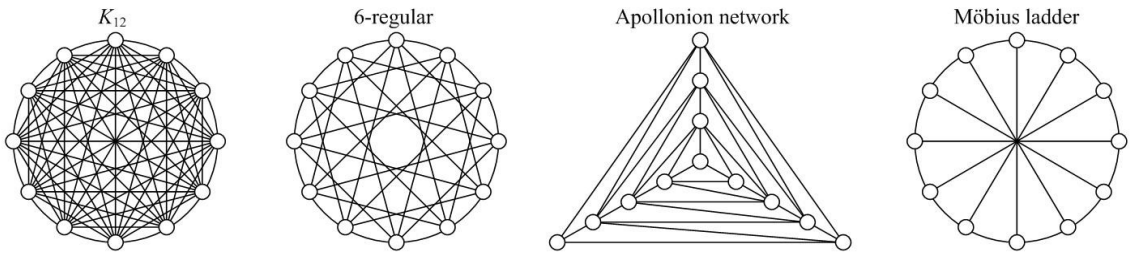

Figure 4. The 12-vertex versions of four of the 48-vertex graphs used as examples above.

A *random graph*, being without an inherent structure, is more malleable than the above graphs (when in a random graph ($V$, $E$) of a given order and size, the members of $E$ have been selected at random from the handshake product of $V$), and a simple way to rearrange such a graph is to order its vertices by degree. Given the graph represented by the table

```
 1 : 2  3  4  5  6  7  8
 2 : 1  4  5  9 11
 3 : 1  8  9 11
 4 : 1  2  6  8  9 11
 5 : 1  2  7  8 11 12
 6 : 1  4  7 10 12
 7 : 1  5  6  8
 8 : 1  3  4  5  7  9 10 12
 9 : 2  3  4  8 11
10 : 6  8 11 12
11 : 2  3  4  5  9 10 12
12 : 5  6  8 10 11
```

the two by-degree arrangements give us two corresponding vertex mappings,

```
 8 : 1  3  4  5  7  9 10 12     8 ->  1        3 : 1  8  9 11                3 ->  1
 1 : 2  3  4  5  6  7  8        1 ->  2        7 : 1  5  6  8                7 ->  2
11 : 2  3  4  5  9 10 12       11 ->  3       10 : 6  8 11 12               10 ->  3
 4 : 1  2  6  8  9 11           4 ->  4        2 : 1  4  5  9 11             2 ->  4
 5 : 1  2  7  8 11 12           5 ->  5        6 : 1  4  7 10 12             6 ->  5
 2 : 1  4  5  9 11              2 ->  6        9 : 2  3  4  8 11             9 ->  6
 6 : 1  4  7 10 12              6 ->  7       12 : 5  6  8 10 11            12 ->  7
 9 : 2  3  4  8 11              9 ->  8        4 : 1  2  6  8  9 11          4 ->  8
12 : 5  6  8 10 11             12 ->  9        5 : 1  2  7  8 11 12          5 ->  9
 3 : 1  8  9 11                 3 -> 10       11 : 2  3  4  5  9 10 12      11 -> 10
 7 : 1  5  6  8                 7 -> 11        1 : 2  3  4  5  6  7  8       1 -> 11
10 : 6  8 11 12                10 -> 12        8 : 1  3  4  5  7  9 10 12    8 -> 12
```

and altogether this gives the three isomorphic graphs shown in Figure 5.

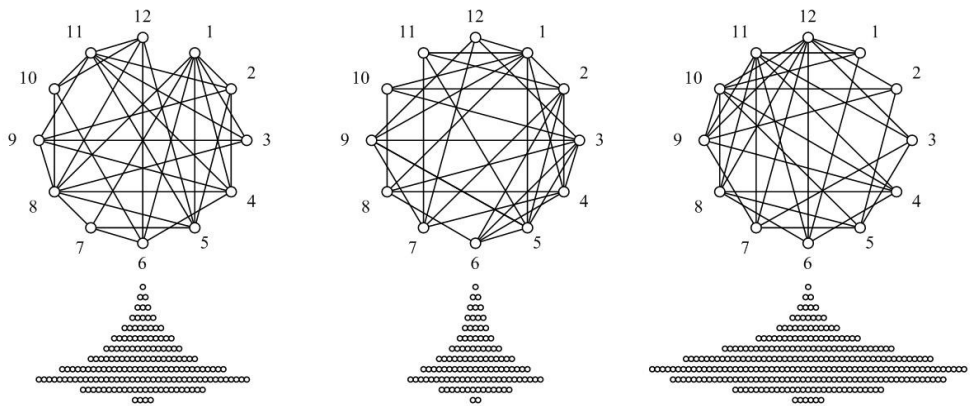

Figure 5. The shapes below the graphs represent the corresponding USBE-trees. As we see, an ordering by descending degree gives a slimmer, and an ordering by ascending degree gives a wider tree than no ordering at all.

In the graph ordered by descending degree in the middle of Figure 5 the densest part contains vertices 1 to 6, which are the first to be removed in an ordered succession of split-by-edge operations, whereas in the graph ordered by ascending degree (to the right) the densest part contains vertices 7 to 12, which are the last to be removed.

*Finding a maximum independent set in a USBE-tree*
Since the maximum independent sets of any graph are the leaves closest to the roots of its USBE-trees, such a set can be found in a layer-by-layer search. We construct each layer $L_i$ from the one above, $L_{i-1}$, in a succession of split operations. In order to avoid duplicates, we use a search tree, and since the sizes of the nodes are specific for each layer, we can use one search tree, $S_i$, per layer. (The fact that $L_i$ and $S_i$ contain the same nodes, seems to indicate that we could have made do with $S_i$ alone, but this has turned out to slow down the search process considerably, possibly as a result of the edge search order having been disrupted.) In a layer-by-layer search, an ordering of the graph's vertices by descending degree more than halves the workload, as illustrated in Figure 6.

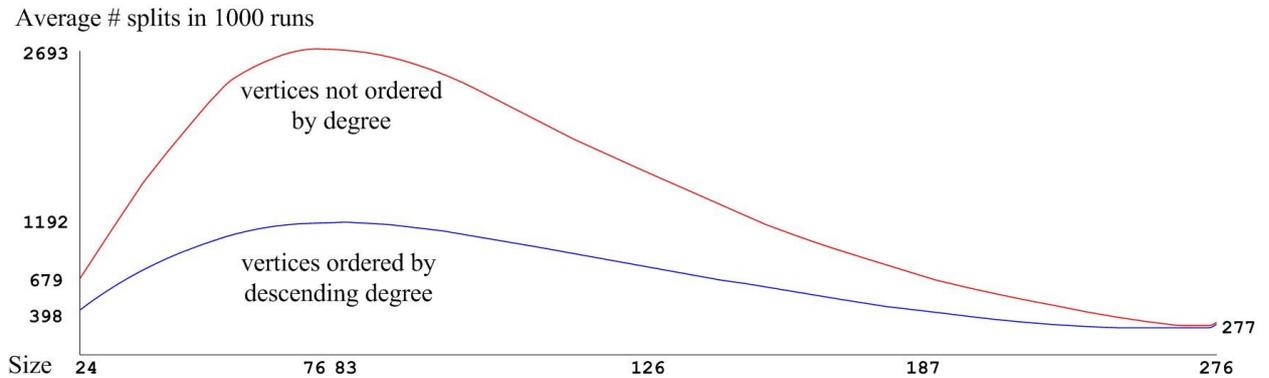

Figure 6. Maximum independent set searches in USBE-trees of random graphs on 24 vertices and *m* edges, for *m* = 24, …, 276, with 1000 runs for each *m*. (That there for some *m* are less than 1000 different graphs, has no bearing on the curve shapes.)

*The complexity of the USBE-tree*
We can describe the complexity of a USBE-tree in terms of the number of split operations, $\delta(n, m)$, required to find a maximum independent set in a graph of order *n* and size *m*. Figure 6 shows the average values of $\delta(24, m)$ in 1000 runs for each *m* from 24 to 276. The maximum of the bottom curve is $\delta(24, 83) = 1192$, which is a little above $2^{0.4n}$. As far as the tests go, the average of the sizes that give the highest split numbers lies a little above $3n$ and the relative maximum number of splits falls from $2^{0.475n}$ to $2^{0.395n}$ for $n = 12$ to $50$.

For comparison, the number of splits for Möbius ladders, for which *n* is always even and $m = 3n/2$, falls from $2^{0.482n}$ to $2^{0.38n}$ in the same interval, and since these are not random we get a definite measure, which is $2^{0.347120956815n + 1.66485616037} - 2$ for *n* divisible by 4, and $2^{0.347120956815n + 1.74055665759} - 2$ for *n* divisible by 2 but not by 4. The values for random graphs are regular enough to indicate that the maximum split number for these are on the same form, $2^{0.369425n + 0.56325}$, to be specific, that is, the maximum number of split operations required to find a maximum independent set in a random graph on *n* vertices is $2^{O(0.369425n)}$.

*Finding and utilising the set of independent sets of a graph*

Let $\mathcal{I}$ be the set of independent sets of a graph $G$, let $T$ be the USBE-tree of $G$, and let $F$ be the foliage of $T$. Then $F \subseteq \mathcal{I}$, but since, by Theorem 1, every set in $\mathcal{I}$ has a superset in $F$, $\mathcal{I}$ can be generated from $F$. In principle, $\mathcal{I}$ is the union of the powersets of the sets in $F$, but rather than generating all these powersets, we iterate over $F^{(i)}$ and $I^{(i)}$ as follows, when $I^{(0)} = F^{(0)} = F$.

Let $$F^{(i)} = \{X \mid \exists Y \in F^{(i-1)}.\ X \subset Y \ \wedge\ |Y| > 1 \ \wedge\ |X| = |Y| - 1 \ \wedge\ X \notin I^{(i-1)}\}$$

and let $$I^{(i)} = I^{(i-1)} \cup F^{(i)}.$$

There must then be a $j$ such that $F^{(j)} \neq \emptyset$ and $F^{(j+1)} = \emptyset$, and then $I^{(j)} = \mathcal{I}$.

$\mathcal{I}$ can be used to find the chromatic number $\chi(G)$ and, knowing $\chi(G)$, to compute all colorings of $G$ (which was what inspired the split-by-edges approach in the first place). We start with the latter.

Knowing $k = \chi(G)$ and $\mu = \alpha(G)$, we compute the set $\mathcal{A}$ of sets of $k$ numbers in $(1, \mu)$ that add up to the order of $G$. We then organize $\mathcal{I}$ into a set of lists $\mathcal{L} = \{L_1, L_2, …, L_\mu\}$ of independent sets of uniform cardinalities, and for each set in $\mathcal{A}$ we find the sets of $k$ non-intersecting sets from the corresponding lists in $\mathcal{L}$, and the union of all of this is then the set of colorings of $G$.

To find $k = \chi(G)$, we use $\mathcal{L}$ as above and iterate over $k$, computing the successive sets of add-up sets $\mathcal{A}_k$, until a coloring is found. For each set $A_j = (a_{j,1}, …, a_{j,k})$ in $\mathcal{A}_k$ we search through the corresponding lists in $\mathcal{L}$, building a set $C$ of non-intersecting independent sets along the way. At some point we have $C = \{C_1, …, C_{i-1}\}$, with one element from each of the respective lists $L_{a_{j,1}}, …, L_{a_{j,i-1}}, i > 1$, and we then search through $L_{a_{j,i}}$.

- A *match*, if any, is a set in $L_{a_{j,i}}$ that does not intersect any of the sets in $C$.
  - If we find a match and $i = k$, we have a $k$-coloring of $G$, and $\chi(G) = k$.
  - If we find a match and $i < k$ we add the match to $C$ and proceed to examine $L_{a_{j,i+1}}$.
  - If we did not find a match we keep looking through $L_{a_{j,i-1}}$ with $C = \{C_1, …, C_{i-2}\}$, or $C = \emptyset$, if $i = 2$.

- If we did not find any match for $A_j$ then
  - if $j < |\mathcal{A}_k|$ we proceed to $A_{j+1}$, and
  - otherwise, we proceed to $k + 1$.

Given a set $\{L_1, …, L_k\}$ of lists of independent sets of uniform cardinalities, if there is a set of sets $\{\{C_1, …, C_k\} \mid C_i \in L_i, i \in (1, k)\}$, this algorithm will give us one of these.

(Of course, operations like these are only practical for relatively small graphs. The graphs of interests are the ones that are too large for naïve coloring algorithms, but small enough for the algorithms described above to work in reasonable time—i.e. seconds and minutes rather than days and weeks.)